\documentclass[aps,preprint,amsfonts,amsmath,amssymb,eqsecnum]{revtex4-1}
\usepackage{verbatim,graphics,graphicx,color,slashed}

\begin{document}


\title{\Large Implications of LHC Searches for Massive Graviton}

\author{ Yong Tang \footnote{ytang@phys.cts.nthu.edu.tw}}
\affiliation{Physics Division, National Center for Theoretical Sciences, Hsinchu, Taiwan}

\date{\today}

\begin{abstract}
With the latest LHC available results, we consider the generic constraints on
massive graviton. Both dijet and dilepton resonance searches are used. The
limits on parameter space can be applied to many models. As an
illustration, we show the constraints for Randall-Sundrum (RS) model.
Implications on massive graviton and the coupling strength are discussed. For
$k/M_{pl}=0.1$, $M_G<2.2$ TeV region is excluded at $95\%$ confidence level. We
also present some interesting implications on the RS radion with respect to the 125 GeV excess at the LHC. For
$k/M_{pl}=0.1$, $\Lambda_\phi<13.8$ TeV is excluded where $\Lambda_\phi$ is the
scale to charactarize the interaction strengh of radion.
\end{abstract}

\maketitle

\section{Introduction}

The standard model (SM) with gauge group SU(3)$\times$SU(2)$\times$U(1)
has been very successful in explanation for experimental results. Although the exact
mechanism for electroweak symmetry breaking is still unclear, the recent hint of
$125$ GeV excess~\cite{ATLAS:2012ae,Chatrchyan:2012tx} may suggest the higgs
mechanism. Then the hierarchy problem in SM, electroweak breaking scale being so
smaller than Planck scale, still exists and motivates new physical ideas. Among
these ideas to solve the hierachy problem,
extra-dimension~\cite{ArkaniHamed:1998rs,Randall:1999ee} is one of the simplest ways .

Numerous models with extra-dimensions describing possible new physics beyond SM
predict massive gravitons~\cite{Giudice:1998ck,Han:1998sg}. The masses of these
particles are usually at the TeV scale if these models were trying to solve the hierarchy problem. Then these
massive gravitons can be produced at the large hadron collider (LHC). In some
cases, the cross section is large enough so that exclusion limit or discovery
can be reached for the considerred models. 

Extra-dimensional models have many phenomenolgical
consequences~\cite{Davoudiasl:2000wi,Csaki:2000zn,Gunion:2003px,Carena:2007ua,Csaki:2008zd,Casagrande:2008hr}.
Before LHC era, flavour physics and electroweak precision observables have already give some limits. In our study,
we shall discuss the constraints from direct searches at the LHC. Other than
specifying on a single model, we will study the general properties of massive
graviton. The discussions and constraints are applicable to a wide class of
models. As an illustration, we show the constraints for Randall-Sundrum (RS) model.

The paper is organized as follows. In section~\ref{sec:Framework}, we introduce
the general framework for later discussion and the necessary ingredients for
searches or exclusions for massive graviton at the LHC. In
section~\ref{sec:constraints}, we use the latest CMS and ATLAS data to constrain
the parameters for the lightest massive graviton. Both dijet and dilepton final
states are considered. In section~\ref{sec:WED}, we show that constraints
has interesting implications in a specific and popular model, warped extra
dimension. Finally, summary and conclusion are given.

\section{Generic framework on massive graviton}
\label{sec:Framework}

Massive gravtions exist in various models usually with extra dimensions. The
compacification of extra-dimensional leads to Kaluza-Klein towers of particle
spectrums. The lowest states of these towers usually are the SM particles, and
the higher states represent these new heavy particles. With exact mass
depending on the details of models, massive Kaluza-Klein particles are roughly
at TeV scale if heirarachy problem is solved. Among them, massive gravitons
serve as indispencible ingredients.

\subsection{Interactions with SM}

The field that describes a spin-2 particle is a tensor, $h_{\mu\nu}$. In a
general effective theory with $h_{\mu\nu}$, all gauge invariant terms, both
renormalizable and non-renormalizable, should be written down in the lagrangian.
However, such a theory will have to too many free parameters to be considered.
Thus, in this paper, we should assume a minimal setup that massive graviton
shall couple to standard model particles with universal form as same as the
massless one, except with different strength and a non-zero mass. This
universality is quite general as long as massive graviton is derived from
the space-time metric. 

For direct production at the LHC, we only need the lowest interaction term,
\begin{equation}
\mathcal{L}_{\rm int}=-\frac{1}{M_{pl}}T^{\alpha\beta}h_{\alpha\beta}^{(0)}
-\frac{1}{\Lambda_{G}}T^{\alpha\beta}h_{\alpha\beta},
\end{equation}
here and after, we will use $h^{0}_{\mu\nu}$ and $h_{\mu\nu}$ for massless
and massive graviton, respectively. $T^{\alpha\beta}$ is the energy-momemtum
tensor of SM field. And the interactions between massive graviton and SM
particle are solely determined by $\Lambda_{G}$. In general, there could be
more than one massive graviton. For simplicity in this work, we will only
concentrate on and refer to the lowest massive graviton, $h_{\mu\nu}$, if not
stated explicitly. Early discussions on searches for massive graviotn are
referred to~\cite{Antoniadis:1999bq,Davoudiasl:1999jd,Accomando:1999sj,Bijnens,
Allanach}.

\subsection{Production at the LHC}

\begin{figure}[t]
\includegraphics[scale=0.5]{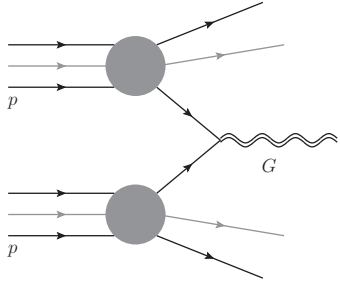}
\caption{Feynman diagram for the massive graviton production at the LHC. The
produced gravitons then decay to SM particles.}
\label{Fig:ProatLHC}
\end{figure}

Based on the interaction term, it is immediately realized that the topology for
production of massive graviton is Drell-Yan like process. At the LHC with
proton-proton collision, the two main channels to produce massive graviton are
gluon-gloun fusion and quark-antiquark annihilation processes, Fig.~(\ref{Fig:ProatLHC}).
Vector boson fusion processes or other assiociated production are neglected for
leading-order approximation throughout this work.

For hadronic collison, the total cross section is the convolution of
the parton distribution fucntions (PDFs) with the partonic cross section, 
\begin{equation}
\sigma=\int
dx_{1}dx_{2}f_{q_{1}}\left(x_{1},\mu_{F}\right)f_{q_{2}}\left(x_{2},\mu_{F}\right)\hat{\sigma}\left(q_{1}q_{2}
\rightarrow G^{*}; \hat{s}\right),
\end{equation}
where $f_{q}(x, \mu_{F})$ is the PDF for a parton $q$(quark and gluon) with
momentum fraction $x$ at the factorization scale $\mu_{F}$, $\hat{\sigma}$ is the partonic cross
section with the initial two partons of momentum fraction $x_1$ and $x_2$,
respectively, and $\hat{s}=x_1x_2s$.

\begin{figure}[t]
\includegraphics[scale=0.28]{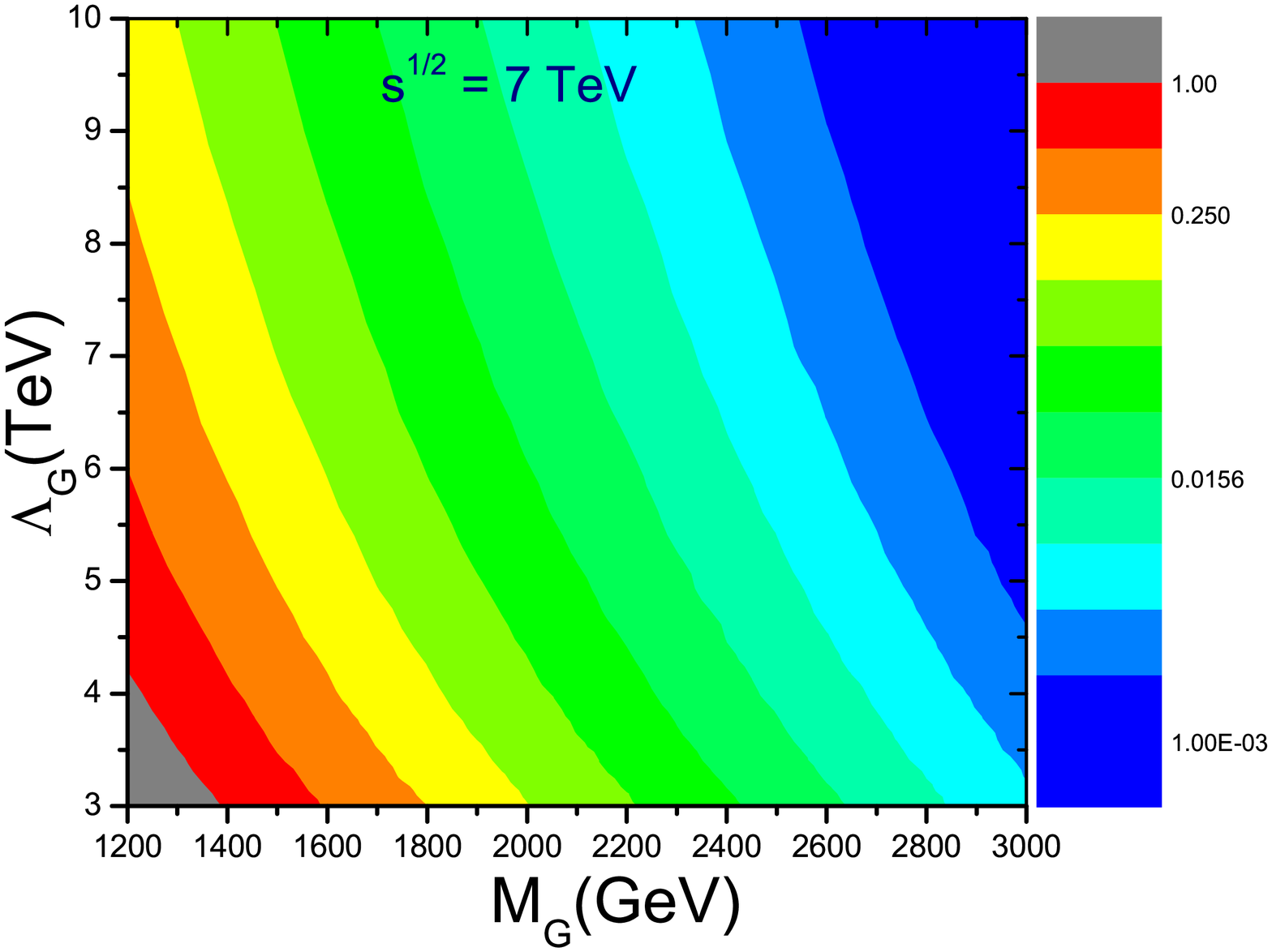}
\includegraphics[scale=0.28]{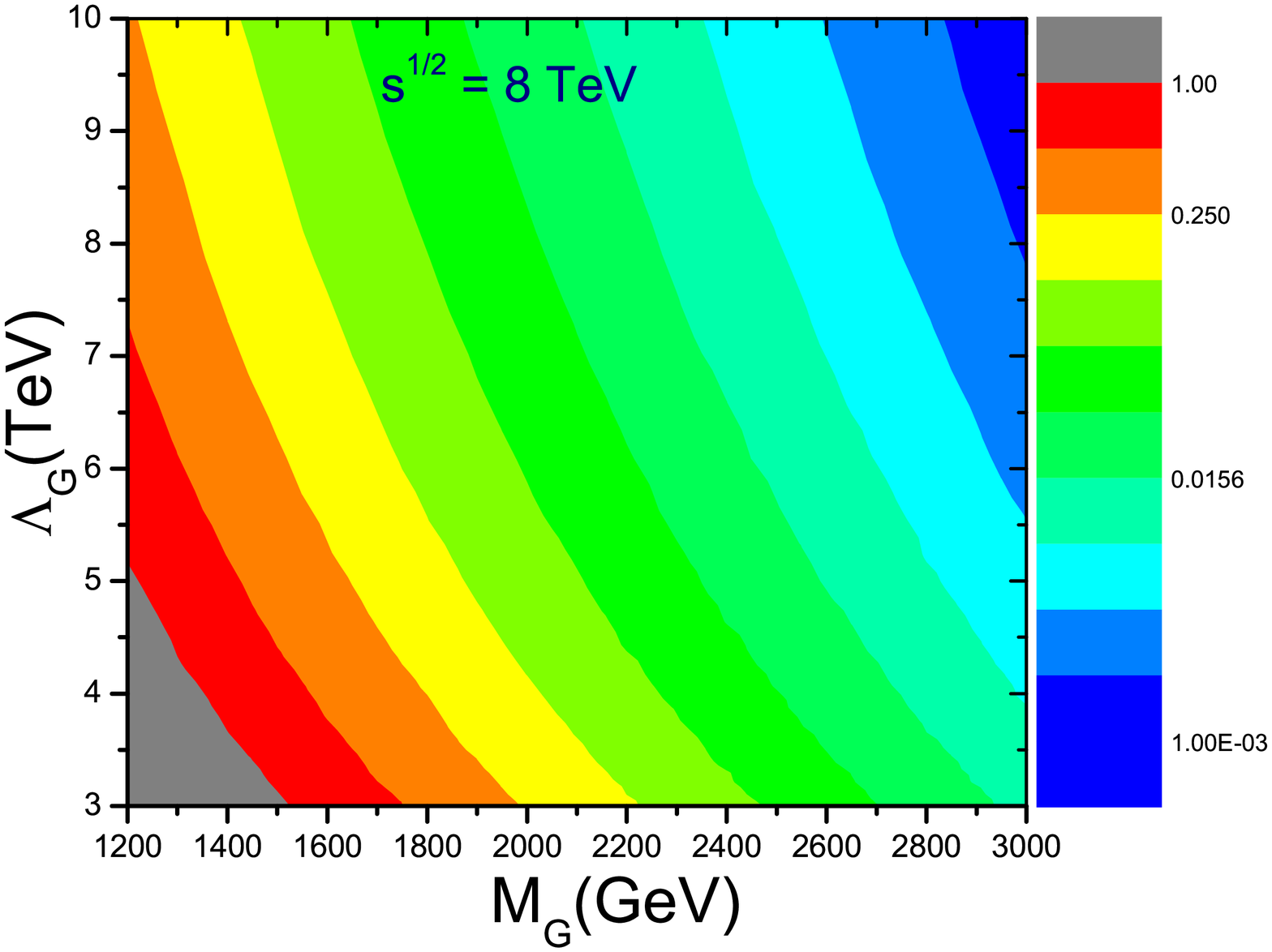}
\caption{Colorred map of cross section (pb) for massive graviton production at
the LHC with $\sqrt{s}=7$ TeV (left) and $\sqrt{s}=8$ TeV (right), respectively.}
\label{Fig:CSatLHC}
\end{figure}

We use Madgraph 5 \cite{Alwall:2011uj, Hagiwara:2008jb} with $\mu_F=M_G$ and
CTEQ6L1 \cite{CTEQ} PDF set, and show in Fig.~(\ref{Fig:CSatLHC}) colorred maps
of cross section with function as mass $M_{G}$ and the scale $\Lambda_{G}$ at
the  LHC with both $\sqrt{s}=7$ TeV (Left figure) and $\sqrt{s}=8$ TeV. The
cross section smaller than 1 {}fb is shown in blue region at the right top
corners.  These two figures roughly show the relations between cross section and
parameters. Compared to the $\sqrt{s}=7$ TeV case, the colored map for
$\sqrt{s}=8$ TeV is shifted towards larger $\Lambda_{G}$ and $M_G$.

\subsection{Decay width and branching ratio}
For direct search or exclusion of the massive graviton, we need to know the
branching ratio of its decay channels. If massive graviton lies in the TeV mass
regions, its decay products, standard model particles, are then on-shell. So
for leading order consideration we only include the two-body decay channels. 
The individual decay rate to two final SM particles is listed in the appendix.
\begin{figure}[tbh]
\includegraphics[scale=0.7]{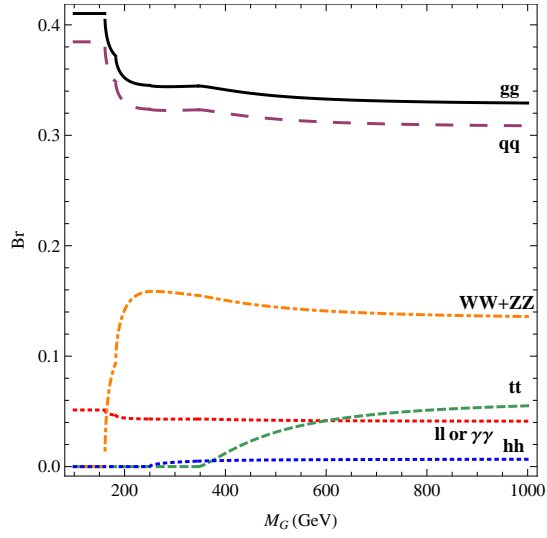}
\caption{Branching ratios, where $q=u,d,c,s,b$ and $l=e,\mu$. We have
$Br(G\rightarrow ll)\simeq Br(G\rightarrow \Gamma\Gamma)$ as shown above.}
\label{Fig:Br}
\end{figure}
In Fig.~\ref{Fig:Br}, we show the branching ratios for the main decay channels.
Here, we neglected the three-body decay with a off-shell $W^*$, $Z^*$ or $t$ at
the low mass range. As shown in the figure, as the mass of graviton goes
large compared with $2m_t$, branching ratios are almost fixed. Now it is easy to get the total decay width
for large $M_{G}$,
\begin{eqnarray*}
\Gamma_G  \simeq 
\frac{M_{G}^{3}}{40\pi\Lambda_{G}^2}\left[(3+1)\times6\times\frac{1}{4}+(1+\frac{1}{2})
\times\frac{13}{12} + (8+1)\times\frac{1}{2}+\frac{1}{12}\right]
  =  \frac{293M_{G}^{3}}{960\pi\Lambda_{G}^2}.
\end{eqnarray*}

In Table.~\ref{Table:Branching}, we show branching ratios of the main decay
channels for later use. The events at partonic level at the LHC then is
proportional ot $\sigma(pp\rightarrow G)\times\mathcal{B}r(G\rightarrow ff)$ for narrow width
approximation.
\begin{table}[b]
\caption{Branching ratio for the massive graviton to SM particle with $l=e,\mu$
and $q=u,d,s,c,b$.}
\begin{tabular}{|c||c|c|c|c|c|c|c|c|}
\hline
\hline
\textrm{channels} & $HH$  & $gg$   & $\gamma\gamma$ & $W^+W^-$ & $ZZ$  &$t\bar{t}$ & $q\bar{q}$ & $ l^+l^- $   \\
\hline
$\mathcal{B}r$    & 2/293 & 96/293 &     12/293     & 24/293   & 12/293&  16/293   &   90/293   & 12/293  \\
\hline
\end{tabular}
\label{Table:Branching}
\end{table}

\section{Constraints from LHC direct searches}
\label{sec:constraints}

The massive graviton couples to standard model particles through the
energy-momentum tensor in the linear theory and the couplings are suppressed by
a factor $\Lambda_G$.  If the mass is at the TeV scale and the factor is not too
large, then enough massive graviton can be produced at the LHC. Searches for
the decay products of the graviton can be used to discover or constrain the parameter space of the
model. Based on the final states, there are various search channels and if
no excess is observed, each channel can give a constraint. In the section, we
focus on the dijet and dilepton final states only. 

\subsection{Dijet Constraint}

Dijets consist of quark jets and gluon jets. The shape of a gluon jet is wider
than that of a quark one because gluon's effective coupling $C_A
\alpha_S$($C_A=3$) is larger than quark's $C_F\alpha_S$($C_F=4/3$), then
gluon jets are more likely to radiate. Modern detectors have the power to
distinguish them. CMS \cite{CMSdijet1fb} has presented limits for three kinds of
dijet resonances, gluon-gluon, quark-quark( quark for $q$ or $\bar{q}$) and
gluon-quark. Because of the larger background from gluon jets, quark-quark dijet
limit is the most stringent one among three .
Then, we shall only consider the constraint from quark-quark dijet spectrum. 

From the branching ratio in Table.~\ref{Table:Branching}, we know that about
$1/3$ of the produced massive graviton will decay to quark-antiquark pairs ($q\bar{q}, q=u,d,s,c,b$).
If the production rate is large, event distribution on dijet invariant mass $m_{jj}$
shall show an additonal peak at the graviton mass in the smooth QCD dijet background. 
If no excess is observed, constraints can be put on production rate using
statistics(see appendix).  So far, the latest published model-independent result
on resonance  search by dijet search is from CMS with 1
fb$^{-1}$\cite{CMSdijet1fb}. Since no excess is observed yet, we shall use this
result  to constrain our parameter space.

The analysis of event samples in \cite{CMSdijet1fb} relies on the following
selection rules,
\begin{equation}
p_{T}>10 \textrm{ GeV},\; |\eta|<2.5,\; |\Delta \eta|<2.5 \textrm{ and } m_{jj}
> 838 \textrm{ GeV},
\end{equation}
where the definitions are
\begin{eqnarray*}
&& m_{jj}\equiv \sqrt{(E_1+E_2)^2-(\vec{p}_1+\vec{p}_2)^2},\;
p_{T}\equiv\sqrt{p^2_x + p^2_y}=p\sin{\theta},\\
&& \eta \equiv -\ln{\tan{\frac{\theta}{2}}},\; \Delta \eta = \eta_1 - \eta_2.
\end{eqnarray*}
The renormalization scale is set to $\mu=p_{T}$ and CTEQ6L1 parton distribution
functions \cite{CTEQ} are used. A K-factor of $1.33$ was used in
\cite{CMSdijet1fb}.

For mass graviton, the signal acceptance $A$ is about $0.72$ for the
events selected that satisfying the above kinematics requirements. This factor
is nearly constant for large mass of $G$. Without seeing any excess of dijet
mass spectrum, an upper limit at the $95\%$ confidence level is obtained for
$\sigma \times Br \times A$, the products of cross section, branching ratio and
events acceptance. Since both $Br$ and $A$ are known, the limit then is
translated to constraint on the $\Lambda_{G}$ and $M_{G}$.

\begin{figure}
\includegraphics[width=0.55\textwidth,height=3.0in]{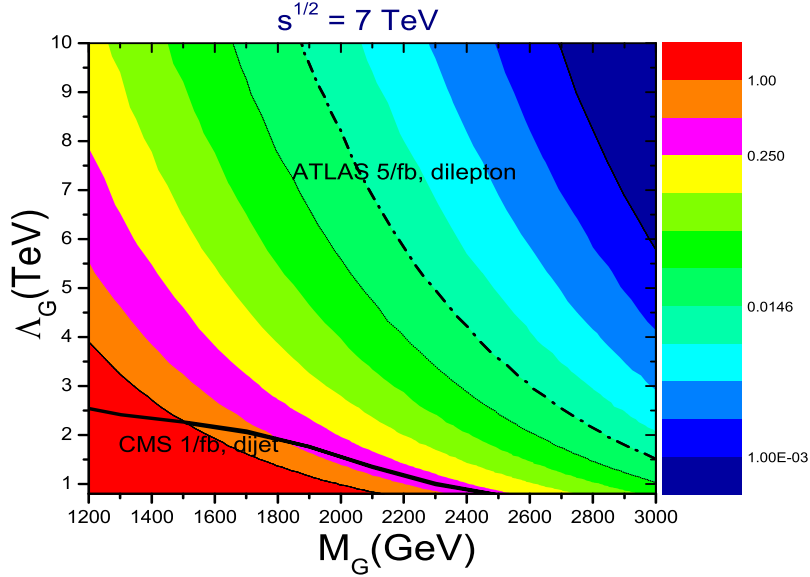}
\caption{No obsearvation of dijet and dilepton event excess can put constraints
on the parameters. Cross section is in pb. The solid line indicates the
constraints from CMS with 1 fb${}^{-1}$ data \cite{CMSdijet1fb} and the dot-dashed line shows the limit put
by ATLAS with 5 fb${}^{-1}$ \cite{ATLASdilepton5fb}.}
\label{fig:Dijet_and_dilepton}
\end{figure}

In Fig.~\ref{fig:Dijet_and_dilepton}, we show the limit as a solid line on the
cross section contour of $\Lambda_{G}$ and $M_{G}$. Regions at the left-handed
side of the line is excluded at $95\%$ confidence level. As is displayed in the
figure, for $M_G$ less than 1 TeV, $\Lambda_{G}$ has to be larger than 2.5 TeV.
As the mass get smaller, $\Lambda_{G}$ needs to larger to accommadate. This is
reasonable and intuitive otherwise we shall have seen a resonance around $M_G$.

\subsection{Dilepton Constraint}
The dilepton searches are involved with dielectron and dimuon final
states where the main SM background is from $Z/\gamma^{\star}$ decay.
For massive graviton, although the branching ratio to
dilepton ($l=e,\mu$) is much smaller than that of dijet (7.5 times smaller
exactly), the constraint or discovery potential is better due to the
well-known and lower background.

The event selection follows~\cite{ATLASdilepton5fb} where for dielectron
\begin{equation}
E_{T}\equiv \sqrt{p_T^2+m^2}>25 \textrm{ GeV},\; |\eta|<2.47 \textrm{ with
excluding } 1.37<|\eta|<|1.52|,
\end{equation} 
and for dimuon $ p_T >25 \textrm{ GeV}.$ The renormalization scale is set to
$\mu=\sqrt{\hat{s}},$ the K-factor for dilepton final state varies between $1.6$ to $1.8$, depending
on the graviton mass and $\Lambda_{G}$. In practice, $1.75$ is used for
$M_G>750$ GeV~\cite{ATLASdilepton5fb}. With these kinematic requirements, the
total signal acceptance $A$ is $72\%$ for dielectron and $47\%$ for dimuon.

As shown in Fig.~\ref{fig:Dijet_and_dilepton} as the dot-dashed line, the
dilepton final state can give more stringent constraint due to both larger
luminosity and better discriminating power, compared with dijet limit with only
1 {}fb data. The limit on cross section is almost constant 2 TeV and 3 TeV
because the constraint mainly comes from the observed
event~\ref{fig:Dijet_and_dilepton} in a single bin $[1200,3000]$ GeV of~\cite{ATLASdilepton5fb}.

\section{Constraint on warped extra dimension}
\label{sec:WED}

\subsection{Theory Overview}

One of the most popular extra-dimensional models is Randall-Sundrum model. The
physics behind RS model lies in the following geometry for the warped
space-time~\cite{Randall:1999ee},
\begin{equation}
ds^2=e^{-2kT(x)|\varphi|}[\eta_{\mu\nu}+G_{\mu\nu}(x)]dx^\mu dx^\nu
+T^2(x)d\varphi^2 ,
\end{equation}
where $T(x)$ is referred to as the modulus field, $G_{\mu\nu}(x)$ as graviton
and $k$ is a scale of the order of the (reduced) Planck scale $M_{pl}$. To
explain the hierarchy problem, the compactification radius or the vacuum
expectation value(vev) of the modulus field, $r_c\equiv\langle T(x)\rangle$, is
required to satisfy the relation $kr_c\sim 12$. 

The action that determines the above geometry is
\begin{equation}
S=-M_{*}^{3}\int d^{5}x\sqrt{g}R^{(5)},
\end{equation}
where the 5D reduced Planck scale $M_{*}$ is related with the 4D (reduced)
Planck scale $M_{pl}$,
\begin{equation}\label{eq:Mrelation}
M_{pl}^{2}=M_{*}^{3}\int_{y=-r_c\pi}^{y=r_c\pi}e^{-2k|y|}dy=\frac{M_{*}^{3}}{k}\left(1-e^{-2kr_c\pi}\right).
\end{equation}

In the original RS model, SM particles are confined to the brane and only
graviton can propogate in the bulk. Then a massive Kaluza-Klein graviton tower
exists besides the massless graviton,
\begin{equation}\label{eq:MGkMpl}
M_{G}^{(n)}=kx_{n}e^{-kr_c\pi}=x_{n}\frac{k}{M_{pl}}\Lambda_G,\;
\Lambda_{G}=M_{pl}e^{-kr_c\pi}, \; J_{1}(x_{n})=0,
\end{equation}
where $J_{1}$ is the Bessel function, and $x_{1}\simeq3.8317,\; x_{2}\simeq7.02,\; x_{3}\simeq10.17,$
and $x_{4}\simeq13.32$. We shall only consider the effect of the lowest
state $M_{G}\equiv M^{1}_G$. Both massless and massive gravitons can couple to
standard model particles,
\begin{equation}
\mathcal{L}=-\frac{1}{M_{pl}}T^{\alpha\beta}h_{\alpha\beta}^{(0)}
-\frac{1}{\Lambda_G}T^{\alpha\beta}\Sigma_{n=1}^{\infty}h_{\alpha\beta}^{(n)}.
\end{equation}
$M_{G}=x_{1}\frac{k}{M_{pl}}\Lambda_G$ implys that the larger $M_{G}$ is, the
larger ratio $k/M_{pl}$ for fixed $\Lambda_{G}$. 

Using the relation between the scale $\Lambda_{G}$ and the mass $M_{G}$, we have
the decay width for massive gravtion in RS model,
\begin{equation}\label{eq:RSwidth}
\Gamma_G  \simeq  \frac{293M_{G}^{3}}{960\pi\Lambda_{G}^2}=\frac{293x_{1}^{2}M_{G}}{960\pi}
 \left(\frac{k}{M_{pl}}\right)^{2} \simeq 1.425M_{G}\left(\frac{k}{M_{pl}}\right)^{2}.
\end{equation}

In the RS original paper, $\frac{k}{M_{pl}}$ was assumed to be less than $1$. 
Most discussions lie in $0.01\le\frac{k}{M_{pl}}\le 1$ for theoretical and experimental studys. The
estimation goes as follows. The 5D curvature scalar is $R_{5}=-20k^{2}$, and
requiring $|R|\simeq M_{*}^{2}$ with Eq.~\ref{eq:Mrelation} gives
\[
20k^{2}\simeq\left(kM_{pl}^{2}\right)^{\frac{2}{3}}\Longrightarrow\frac{k}{M_{pl}}
\simeq\left(\frac{1}{20}\right)^{\frac{3}{4}}\simeq 0.1  .
\]
However, it was argued in \cite{Luty,Agashe:2007zd,Grzadkowski:2012ng} that
$R_5$ should be compared with $\Lambda^{2}$($\Lambda \equiv 24^{1/3}\pi M_{*}$ is the energy scale at which
the 5D gravity theory becomes strongly coupled), giving
\[
20k^{2}<\Lambda^{2}\Longrightarrow\frac{k}{M_{pl}}<\left(24^{2/3}\pi^{2}\times
\frac{1}{20}\right)^{\frac{3}{4}}\sim2.88 .
\]

When $k/M_{pl}$ is large, the decay width in Eq.~\ref{eq:RSwidth} shows that
it can be comparable with or even large than its mass. This can happen for
strong interaction. For example, in PDG~\cite{Nakamura:2010zzi} 
$m_{f_{0}\left(600\right)}=\left(400-1200\right)$ MeV, but its width is in the
range, $\Gamma=\left(600-1000\right)$ MeV. And $m_{\rho\left(770\right)}=775.49$
MeV with width $\Gamma=149.1$ MeV~\cite{HYCheng}. The finite decay width effect
will be considered in the following section.

\subsection{Limits on RS graviton}

\begin{figure}[tbh]
\includegraphics[width=0.45\textwidth,height=2.3in]{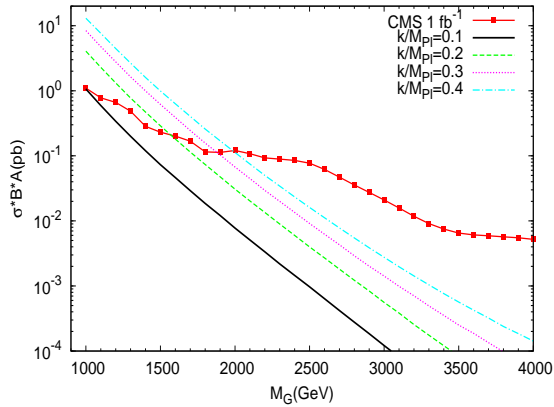}
\caption{Constraint on the parameters $M_G$ and $\Lambda_G$ with dijet events at
CMS 1~fb \cite{CMSdijet1fb}.}
\label{fig:Gjj_1}
\end{figure}

\begin{figure}[tbh]
\includegraphics[width=0.45\textwidth,height=2.3in]{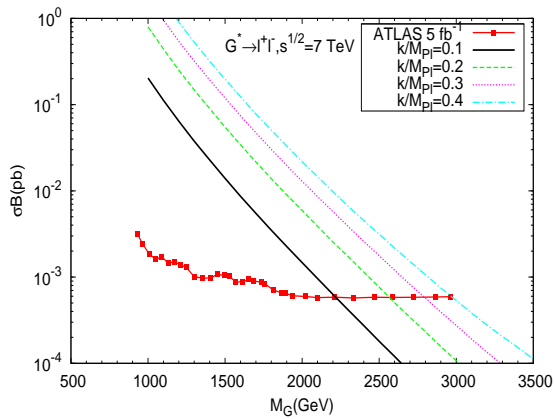}
\caption{Constraint on the parameters with dilepton events $M_G$ and
$\Lambda_G$, for various $k/M_{pl}$. The box points are extracted from
\cite{ATLASdilepton5fb}.}
\label{fig:Gll_1}
\end{figure}

For the lightest massive graviton in RS model, we have the relation
Eq.~\ref{eq:MGkMpl}, $M_{G}=x_1\frac{k}{M_{pl}}\Lambda_G$. This feature shows
that for fixed $M_G$, $k/M_{pl}$ can effectively discribe the interaction
strength. Larger $k/M_{pl}$ means smaller $\Lambda_G$ and then stronger
interaction. Both Tevatron and LHC has set some exclusion limit for
$0.01\leq \frac{k}{M_{pl}}\leq 0.1$. From the latest dilepton search result
\cite{ATLASdilepton5fb}, the $k/M_{pl}=0.1$ gives $M_{G}>2.16$ TeV,
implying $\Lambda_{G}>5.64$ TeV. 

In the following discussion, we shall extend the limit to $k/M_{pl}\geq 0.1$
region. As shown in Fig.~\ref{fig:Gjj_1} and \ref{fig:Gll_1}, both the dijet and
dilepton exclusion limits depend on the $k/M_{pl}$. In the dijet case,
$M_G\leq 1$ TeV is excluded for $k/M_{pl}=0.1$ and the limit turns higher
mass for larger $k/M_{pl}$. In the dilepton figure, $M_G\leq 2.2$ TeV is
excluded for $k/M_{pl}=0.1$, and again larger $k/M_{pl}$ gives even more
stringent constraints. 

For Drell-Yan like s-channel production followed by immediate decay, there is a
propogator of Breit-Wigner like
\[
\frac{1}{\hat{s}-M_{G}^{2}+iM_{G}\Gamma_{G}},
\]
where the width of the massive particle $\Gamma_{G}$ has been included. Usually,
the width is very small for weak interactions, the narrow width approximation
can be used and the final cross section is product of cross sction for
massive particle and branching ratio to final states. For large decay width, the
above full Breit-Wigner propogator is needed. 

\begin{figure}[tbh]
\includegraphics[width=0.45\textwidth,height=2.3in]{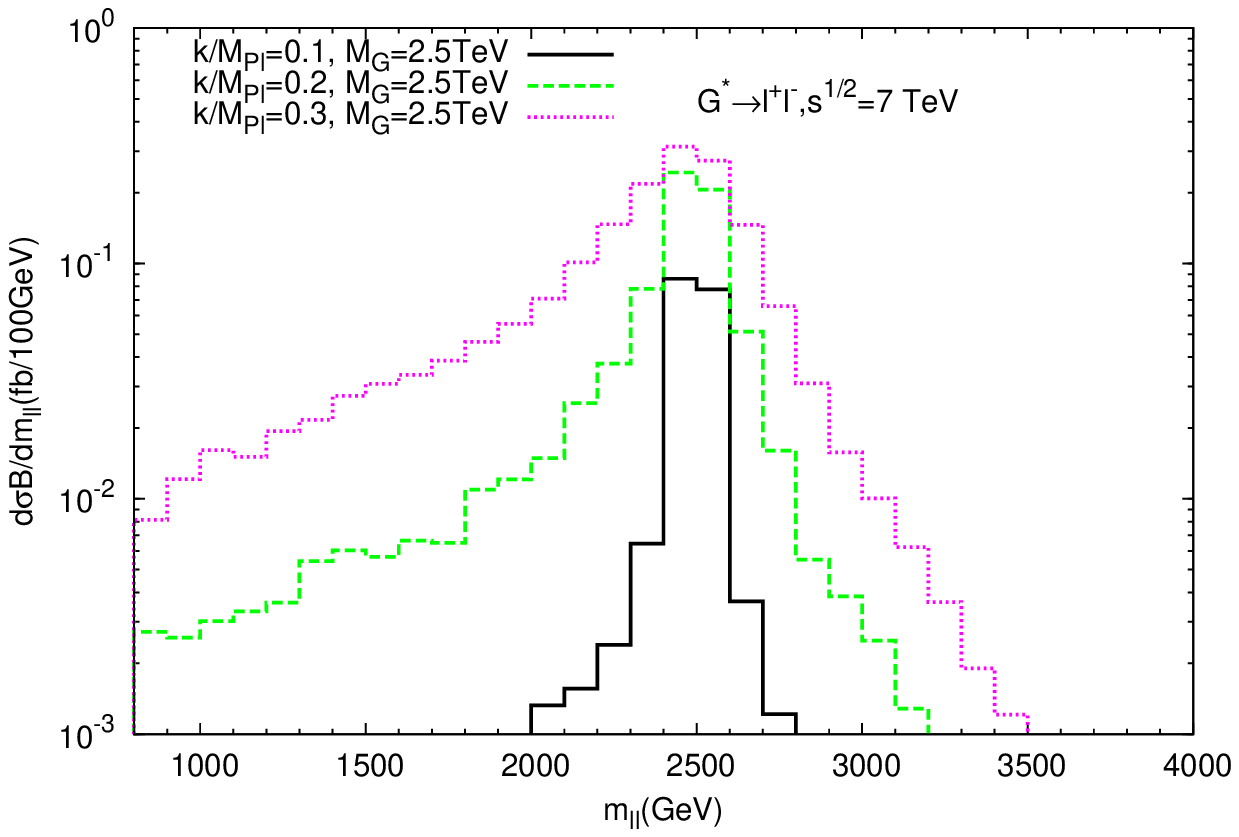}
\includegraphics[width=0.45\textwidth,height=2.3in]{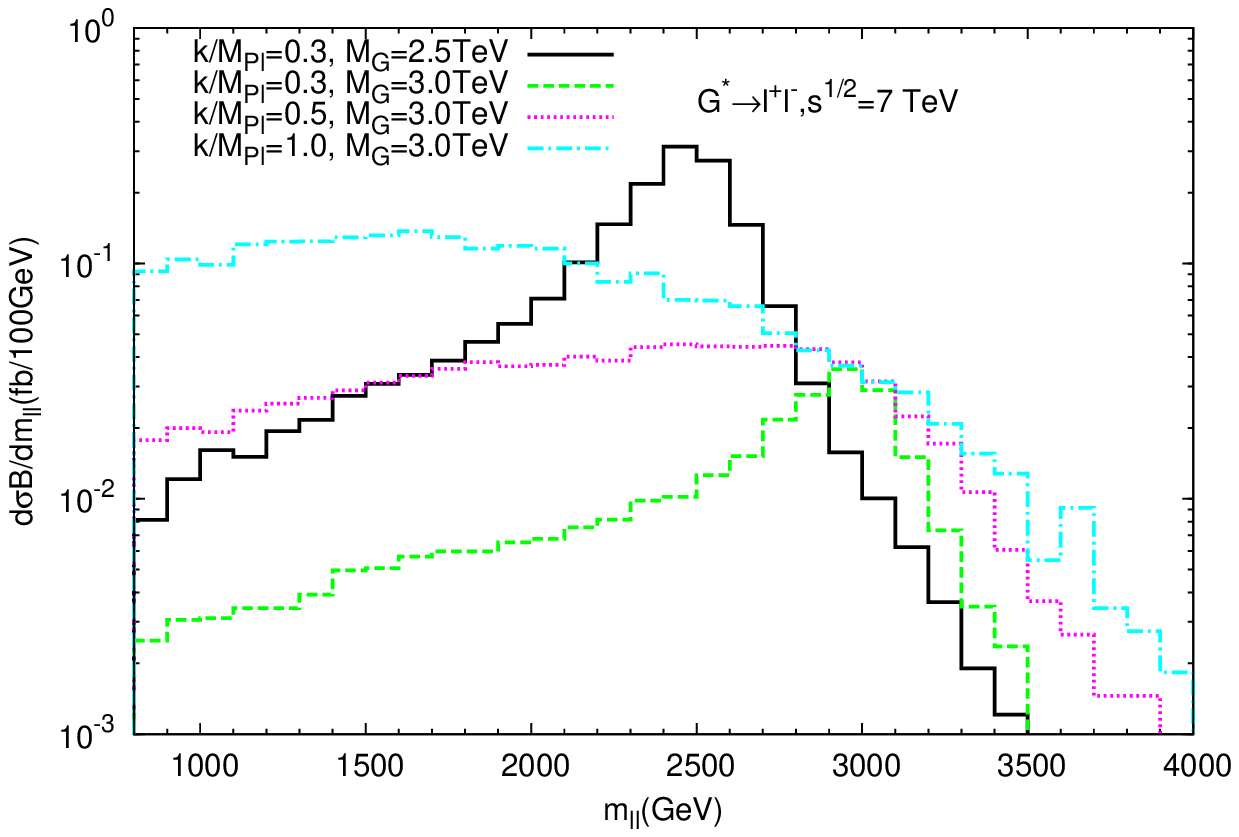}
\caption{Differential cross section as distributions with invarirant mass
$m_{jj}$. Two cases with $M_G=2.5$~TeV(left) and $M_G=3$~TeV(right) are shown. }
\label{fig:Gll_2}
\end{figure}

How the events are distrubted with respect to the $m_{ll}$ depends on PDfs
$f_{q}\left(x,\mu_{F}\right)$, $M_G$ and $k/M_{pl}$. In the
Fig.~\ref{fig:Gll_2}, we show several cases with different $M_G$ and
$k/M_{pl}$. When $k/M_{pl}<0.3$, a clear resonance is still visible due to
$\Gamma_G\leq 0.2M_{G}$ for $k/M_{pl}\leq 0.38$. When $k/M_{pl}$ is large, the
decay width goes large and the resonance get broadened.

\begin{figure}[t]
\includegraphics[width=0.8\textwidth,height=3.0in]{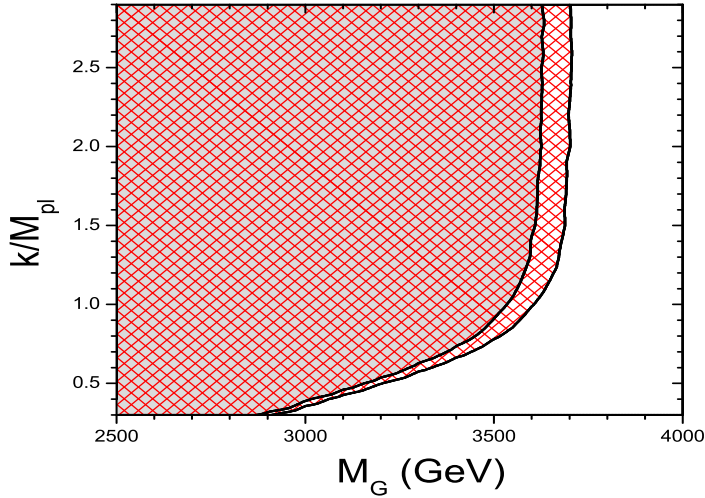}
\caption{Conservative limit on $k/M_{pl}$ and $M_G$, where the shadowed region
is excluded at least at $95\%$ confidence level and the region between two solid
lines indicates the effect of $10\%$ uncertainty.}
\label{fig:Contour}
\end{figure}

The limits are shown in Fig.~\ref{fig:Contour}. For $M_G>3.5$ TeV, the
constraint is allmost fixed for large $k/M_{pl}$. This is simply due to $k/M_{pl}$ cancellation between the
coupling and decay width. And the processes can be discribed by a four-fermion
contact interaction.


\subsection{Implications for RS Radion}

Randall-Sundrum scenario \cite{Randall:1999ee} is proposed to solve the
hierarchy problem of the standard model. In the original version, no mechanism for dynamical
origin of the $kr_{c}\pi$ is provided. Later, Goldberger and
Wise \cite{Goldberger:1999uk,Goldberger:1999un} introduced a mechanism to
provide a potential for radion to stabilize the extra dimension. The radion,
identitied as the gravitational degree of freedom reponsible for the
fluctuations of the distance between branes, couples with standard model particles with similarity of higgs
boson~\cite{Barger:2011qn,deSandes:2011zs,Barger:2011hu}, except that couplings
with massless gauge bosons could be larger.

The indications of 125 GeV excess at the LHC has intrigued many interesting
discussions related with
radion~\cite{Cheung:2011nv,Grzadkowski:2012ng,Frank:2012nb,Tang:2012uw,Davoudiasl:2012xd,Chang:2012tb}.
As shown in \cite{Cheung:2011nv}, the excess observed at the LHC can be
explained by a $125$ GeV RS radion with $\sigma(H)Br(H\rightarrow
\gamma\gamma)/\sigma Br_{\textrm{SM}}\sim 2.1$ and smaller values for other
channels relative to the corresponding ones in SM. Further phenomological
discussions are showed
in~\cite{Grzadkowski:2012ng,Frank:2012nb,Tang:2012uw,Davoudiasl:2012xd,Chang:2012tb}.

The radion $\phi$ couples to SM particles
\cite{Goldberger:1999uk,Goldberger:1999un} as
\[ 
\mathcal{L}_{int} =\frac{\phi}{\Lambda_\phi}T^\mu{}_{\mu},
\]
where $T_{\mu\nu}$ is energy-momentum tensor for SM particles and $\Lambda_\phi
= \sqrt{6}M_{pl}e^{-kr_c\pi}$. Thanks to the trace anamoly, this model leads to
a larger branching ratio for $\phi\rightarrow gg \textrm{ or } \gamma\gamma$, relative to
$h_{\textrm{SM}}\rightarrow gg \textrm{ or } \gamma\gamma$ in SM.

In this section, we show that the results of LHC searches for massive
graviton have several interesting implications for the radion sector in RS
model. The first and lightest massive Kaluza-Klein(KK) mode of $G_{\mu\nu}$ will
couple to SM particles as $
\mathcal{L}_{int}=-\frac{1}{\Lambda_{G}}h_{\mu\nu}T^{\mu\nu} $.
As is shown above, the couplings of the massive graviton with SM particles are
proportional to $1/\Lambda_{G}$ or $x_1k/M_{pl}$ for a fixed $M_G$. Limits put
on $M_{G}$ for specified $k/M_{pl}$ can then be translated to limits on
$\Lambda_{G}$, therefore constraints on $\Lambda_{\phi}$ due to the relation
$\Lambda_{\phi}=\sqrt{6}\Lambda_{G}$.

Using dijet final states from CMS \cite{CMSdijet1fb} with $1 \textrm{
fb}^{-1}$ data, we can exclude a RS graviton mass below $1$ TeV for
$k/M_{pl}=0.1$ in Fig.~\ref{fig:Gjj_1}. A straightforward calculation gives 
$\Lambda_{\phi}=\frac{\sqrt{6}}{x_1 k/M_{pl}}M_G=6.4$ TeV.
With dilepton final states from ATLAS \cite{ATLASdilepton5fb} with $5
\textrm{ fb}^{-1}$, we show in Fig.~\ref{fig:Gll_1} that a RS graviton mass
below $2.2$ TeV is excluded at $95\%$ confidence level with $k/M_{pl}=0.1$, then the corresponding
$\Lambda_{\phi}\simeq 13.8$ TeV.

A smaller value of $\Lambda_{\phi}$ then requires a larger $k/M_{pl}$, although
the latter of order $0.1$ or less is preferred theoretically
\cite{Davoudiasl:2000wi}. However, a larger $k/M_{pl}$ means a more stringent
constraint on $M_{G}$ because the cross section for the graviton's production at
the LHC is proportional to $(k/M_{pl})^2$. As shown in Fig.~\ref{fig:Contour},
when $k/M_{pl}=0.3$, the limit for $M_G$ is $2.8$ TeV, then we have $\Lambda_{\phi}=5.97$ TeV.
A limit of $M_G=3.5$ TeV will give $\Lambda_{\phi}=2.24$ TeV for $k/M_{pl}\simeq
1$. Even the largest but highly theoretically disfavoured $k/M_{pl}\simeq 2.88$ results in
$\Lambda_{\phi}=0.8$ TeV and $\sigma(H)Br(H\rightarrow \gamma\gamma)/\sigma Br_{\textrm{SM}}\sim 1.5$.

\section{Summary}
In this work, we discuss the constraints on massive graviton with the latest LHC
data. In a general framework, we show both dijet and dilepton limits on the two
parameters, the mass $M_G$ and coupling strengh $\Lambda_G$. The limits are
applicable to a wide class of models. As an illustration, we discuss the
implications for RS massive graviton. For $k/M_{pl}=0.1$, the dilepton search at
ATLAS with 5 fb${}^{-1}$ excluded $M_G<2.2$ TeV regions. The constraint becomes
more stringent for larger $k/M_{pl}$.

As a byproduct, we show in RS model that constraints on massive graviton can
give interesting implications on the radion sector. For $k/M_{pl}=0.1$, the dijet
search at CMS with 1 fb${}^{-1}$ excluded $\lambda_\phi<6.4$ TeV intervals.
And the dilepton search at ATLAS with 5 fb${}^{-1}$ can exclude
$\lambda_\phi<13.8$ TeV regions. For $k/M_{pl}\simeq 1$, the low limit get relaxed to
$\lambda_\phi\simeq 2.24$ TeV. These limits have a direct impact on the cross
secction $\sigma(pp\rightarrow \phi)$ of radion production at the LHC since
$\sigma(pp\rightarrow \phi)\propto 1/\Lambda^2_{\phi}$.

\section*{Acknowledgments}
The author would like to thank Prof.~H.~Y.~Cheng, K.~Cheung, K.~Hagiwara, and
C.~Q.~Geng for helpful conversations. This work is supported by National
Center for Theoretical Sciences, Hsinchu.

\section{Appendix}
\subsection{Decay rate}
This subsection lists the decay rates for graviton to two standard model
particles. The complete descriptions and full Feynman rules are refered to
\cite{Giudice:1998ck,Han:1998sg}. The decay rate to two fermions, $G\rightarrow
f\bar{f}$, is
\begin{equation*}
\Gamma \left(G \rightarrow f\bar{f} \right) =
N_{c}\frac{M_{G}^{3}}{160\pi\Lambda_{G}^2} \left(1-4x_{f}\right)^{\frac{3}{2}}
\left(1+\frac{8}{3}x_{f}\right),
\end{equation*}
where $N_c$ is equal to 3 for quarks and 1 for leptons. For vector weak
bosons final states, the rate is 
\begin{equation*}
\Gamma\left(G\rightarrow WW/ZZ\right)=\delta\frac{M_{G}^{3}}{40\pi\Lambda_{G}^2}
\left(1-4x_{V}\right)^{\frac{1}{2}}\left(\frac{13}{12}+\frac{14}{3}x_{V}+4x_{V}^{2}\right),
\end{equation*}
where $\delta=1(\frac{1}{2})\textrm{ for }W(Z)$, respectively. For the
massless final states, gluon and photon, we have
\begin{equation*}
\Gamma\left(G\rightarrow gg/\gamma\gamma\right)=N_{G}\frac{M_{G}^{3}}{80
\pi\Lambda_{G}^2}.
\end{equation*}
Here $N_{G}=1(8)\textrm{ for }\gamma(g)$. Finally, the decay rate to higgs
bosons is
\begin{equation*}
\Gamma\left(G\rightarrow
HH\right)=\frac{M_{G}^{3}}{480\pi\Lambda_{G}^2}\left(1-4x_{H}\right)^{\frac{1}{2}}.
\end{equation*}
In all the above formulas, $x_i=m_{i}^{2}/M_{G}^{2}, i=f,V,H$.

\subsection{Bayesian inference}
In this subsection, we briefly outline the statistics used the the context.
Based on Bayesian inference, given the data or events observed, one can estimate
the probability density of the parameter $s$ (the signal or cross section, for
example) predicted by a theoretical model,
\begin{equation*}
p(s|{\rm data})=\frac{\mathcal{L}({\rm data}|s)\pi(s)}{\mathcal{N}},
\end{equation*}
$p(s|{\rm data})$ is the posterior probability density function(PDF),
$\mathcal{L}({\rm data}|b,s)$ is the likelihood function, $\pi(s)$ is the prior
PDF, and $\mathcal{N}$ is the normalization constant,
\begin{equation*}
\mathcal{N}=p({\rm data}|s)=\int \mathcal{L}({\rm data}|s) {\rm d} s.
\end{equation*}
Upper limit $x$ can be put on $s$ with $95\%$ confidence level when
\begin{equation*}
\int^{x}_{0}p(s|{\rm data}){\rm d}s=0.95.
\end{equation*}
In a counting experiment, events are the sum of background and singal, the
likelihood function is actually $\mathcal{L}({\rm data}|b,s)$ which is given by
\begin{equation*}
\mathcal{L}({\rm data}|b,s)={\rm Poisson}(d|b,s)=\frac{(b+s)^{d}}{d!}e^{-(b+s)},
\end{equation*}
where $d$ is the number of the observed events, $b$ and $s$ are predicted by 
theory for background and signal, respectively. For example, when $d=2$,
$b=0.92$, then the above formalism give the upper limit with $95\%$
confidence level $s \leq x = 5.45$. Since $s$ is proportional to
the cross section which is a function of the parameters in the model. Then
limit set on $s$ can be translated to the parameters.

For data collected into $N$ bins, the above formulas can be generalized to
\begin{equation*}
\mathcal{L}({\rm data}|b,s)=\prod_{i=1}^{N}{\rm Poisson}(d_i|b_i,s_i)
= \prod_{i=1}^{N}\frac{(b_i+s_i)^{d_i}}{d_i!}e^{-(b_i+s_i)},
\end{equation*}
where $d_i,b_i,s_i$ corresponds the quantities in the $i$-th bin. Details and
examples can be found in \cite{Choudalakis:2011bf} where uncertainty effect is
also discussed.

\end{document}